\documentclass
[%
reprint,
superscriptaddress,
 amsmath,amssymb,
 aps,
pra,
]{revtex4-2}

\usepackage{subfigure}
\usepackage{graphicx}% Include figure files
\usepackage{dcolumn}% align table columns on decimal point
\usepackage{bm}% bold math
\usepackage{xcolor}

\usepackage[mathlines]{lineno}% Enable numbering of text and display math
\modulolinenumbers[5]% Line numbers with a gap of 5 lines
\begin{document}

\preprint{aaPM/123-QED}

\title[Sample title]{Theoretical analysis of a Polarized Two-Photon Michelson Interferometer with Broadband Chaotic Light }% Force line breaks with \\
%\thanks{Footnote to title of article.}

\author{Yu Zhou}
%\email{Equal contribution}
\affiliation{MOE Key Laboratory for Nonequilibrium Synthesis and Modulation of Condensed Matter, Department of Applied Physics, Xi'an Jiaotong University, Xi'an, Shaanxi 710049, China
}
\affiliation{Electronic Materials Research Laboratory, Key Laboratory of the Ministry of Education $\&$ International Center for Dielectric Research, School of Electronic Science and Engineering, Xi'an Jiaotong University, Xi'an, Shaanxi 710049, China}%

\author{Sheng Luo}
%\email{Equal contribution}
%\footnote{zhou1@mail.xjtu.edu.cn}
\affiliation{MOE Key Laboratory for Nonequilibrium Synthesis and Modulation of Condensed Matter, Department of Applied Physics, Xi'an Jiaotong University, Xi'an, Shaanxi 710049, China
}
\affiliation{Electronic Materials Research Laboratory, Key Laboratory of the Ministry of Education $\&$ International Center for Dielectric Research, School of Electronic Science and Engineering, Xi'an Jiaotong University, Xi'an, Shaanxi 710049, China}%

\author{Jianbin Liu}
\email{liujianbin@mail.xjtu.edu.cn}
\affiliation{Electronic Materials Research Laboratory, Key Laboratory of the Ministry of Education $\&$ International Center for Dielectric Research, School of Electronic Science and Engineering, Xi'an Jiaotong University, Xi'an, Shaanxi 710049, China}%

\author{Huaibin Zheng}
\email{huaibinzheng@mail.xjtu.edu.cn}
\affiliation{Electronic Materials Research Laboratory, Key Laboratory of the Ministry of Education $\&$ International Center for Dielectric Research, School of Electronic Science and Engineering, Xi'an Jiaotong University, Xi'an, Shaanxi 710049, China}%

\author{Hui Chen}
\affiliation{Electronic Materials Research Laboratory, Key Laboratory of the Ministry of Education $\&$ International Center for Dielectric Research, School of Electronic Science and Engineering, Xi'an Jiaotong University, Xi'an, Shaanxi 710049, China}%
\author{Yuchen He}
\affiliation{Electronic Materials Research Laboratory, Key Laboratory of the Ministry of Education $\&$ International Center for Dielectric Research, School of Electronic Science and Engineering, Xi'an Jiaotong University, Xi'an, Shaanxi 710049, China}%
\author{Yanyan Liu}
\affiliation{Science  and  Technology  on Electro-Optical  Information  Security Control  Laboratory, Tianjin 300308, China}%

\author{Fuli Li}
\affiliation{MOE Key Laboratory for Nonequilibrium Synthesis and Modulation of Condensed Matter, Department of Applied Physics, Xi'an Jiaotong University, Xi'an, Shaanxi 710049, China
}
\author{Zhuo Xu}
\affiliation{Electronic Materials Research Laboratory, Key Laboratory of the Ministry of Education $\&$ International Center for Dielectric Research, School of Electronic Science and Engineering, Xi'an Jiaotong University, Xi'an, Shaanxi 710049, China}%

\date{\today}% It is always \today, today,

\begin{abstract}
In this paper, we study two-photon interference of broadband chaotic light in a Michelson interferometer with two-photon-absorption detector. The theoretical analysis is based on two-photon interference  and Feynman path integral theory. The two-photon coherence matrix is introduced to calculate the second-order interference pattern with polarizations being taken into account. Our study shows that the polarization is another dimension, as well as time and space, to tune the interference pattern in the two-photon interference process. It can act as a switch to manipulate the interference process and  open the gate to many new experimental schemes.
\end{abstract}

%Use showkeys class option if keyword
\maketitle

\section{\label{sec:leve$L_1$}Introduction}
Michelson Interferometer (MI), as an important instrument to study the temporal coherence of electromagnetic (EM) fields, has been applied to many important scientific research projects including the well known Laser Interferometer Gravitational-wave Observatory (LIGO)  \cite{Harry2010Advanced}. A two-photon absorption(TPA) detector can be triggered by a pair of photons when the difference of their arriving time is in the  order of a few femtoseconds  \cite{2002Ultrasensitive,1998Generation,1998Ultrahigh}. The combination of a MI with a TPA detector is used to study the Hanbury Brown
and Twiss (HBT) effect of chaotic thermal light. For ordinary detectors the coherence time of chaotic light which is at the order of femtoseconds is too short. The MI provides the interference paths and TPA detector responses in ultra-short coherence time. Many state-of-the-art researches has been done with the setups, such as measuring photon bunching effect of real chaotic light from a black body \cite{boitier2009measuring}, observing the interference between photon pairs from independent chaotic sources \cite{2011Indistinguishable}, finding the polarization time of unpolarized light \cite{shevchenko2017polarization} \textit{etc}. The similar setup has also been used to recover the hidden polarization \cite{2018Recovering} and form ultra-broadband ghost imaging \cite{2015Ultrabroadband} \textit{etc}. Instead of chaotic sources, the quantum light source like entangled photon pairs and ultra-bright twin beams has also been studied by using this kind of setups \cite{2012Coherence,boitier2011photon}.  

In Ref.\cite{tang2018measuring} the super-bunching effect of photons of true chaotic light  was experimentally demonstrated in the similar setup by cascading the interferometer. Moreover, we proposed to explore the super-bunching effect to enhance the sensitivity of weak signal (such as gravitational wave) detection. To do so it is critical to manipulate the two-photon interference in the setup to increase the interference effect. According to previous studies, we realized that polarization is a parameter as same as space and time in the two-photon interference phenomenon. It could help us to manipulate the two-photon interference in a MI. A theory based on two-photon interference and Feynman path integral which also taking polarization into consideration is necessary for future research. However, the two-photon interference theory reported in previous publication does not take polarizations of $EM$ fields into considerations \cite{2012Coherence,tang2018measuring}. Some of the previous studies on polarization in a MI are from angle of classical coherence theory \cite{2008Polarization,2019Interference}.

Therefore in this paper we analyze a polarized MI with broadband chaotic light detected by a TPA detector with quantum theory.  The theoretical model is based on quantum two-photon interference and Feynman path integral theory. In the analysis we expand the scalar model \cite{tang2018measuring} to vector model by taking polarizations into consideration and introduce a two-photon covariance matrix to describe the transformation of two-photon coherence in the MI. We analyze the four components of the TPA detection in the scalar model and connect them with interference between different two-photon probability amplitudes. It is found that in the vector model polarizations work as a switch to control the coefficients of the four components of TPA detection output where in a scalar model the coefficients are all equal. By adjusting the polarizers in the MI we can make some component to be zero or dominating. For example, we can choose to observe only the HBT effect (with constant background), observe sub-wavelength effect by removing $\omega$ oscillation component, or make $\omega$ oscillation component dominate over $2\omega$ oscillation component \textit{etc}. The model suggests new experimental schemes. It can also help us to further study the manipulation of two-photon interference to explore super-bunching effect in weak signal detection \cite{tang2018measuring}. This model can also be applied to study the MI with polarized quantum sources such as entangled photon pairs or squeezed light \textit{etc}.

\section{\label{sec:theory}Theory}

\begin{figure}
\centering
\hspace*{-.5cm}
\includegraphics[totalheight=2.4in]{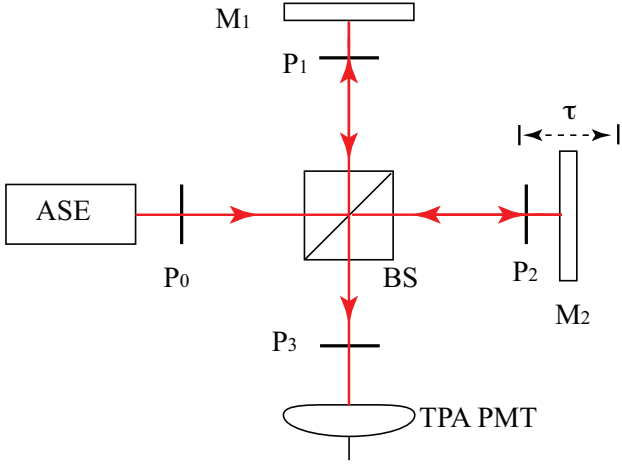}% Here is how to import EPS art
\dimendef\prevdepth=0
\caption{\label{fig:epsart} A Michelson interferometer with polarizers. The polarizers $P_0$ can change the unpolarized chaotic light into linear polarized light. $P_1$ and $P_2$ are in arms $1$ and $2$ respectively. $P_3$ is in front of the detector and can erased the \textit{which path} information. The mirror $M_1$ is fixed and mirror $M_2$ can scan in the longitudinal direction.}
\label{fig:setup}
\end{figure}

The HBT effect can be described as the results of interference between two different but indistinguishable two-photon probability amplitudes \cite{fano1961quantum}. The interfering phenomenon in a MI with broadband  chaotic light detected by a TPA detector can also be understood in the same way. The detection scheme is shown in Fig.~\ref{fig:setup}. 

A continuous amplified spontaneous emission (ASE) incoherent light is used in the configuration. The ASE is completely unpolarized light just like natural light\cite{boitier2009measuring}. The wavelength of the ASE is center at $1550 nm$ with $30 nm$ bandwidth. ASE is coupled into the MI which consists of two mirrors ($M_1$ and $M_2$) and a beam splitter (BS). There are four polarizers $P_0$, $P_1$, $P_2$ and $P_3$ could be put in or taken away from the MI depends on different experiments. $P_0$ could be put at the input of the interferometer . $P_1$ and $P_2$ could be put at two arms of the MI in front of mirrors $M_1$ and $M_2$ respectively. The output beam of the interferometer goes into a semiconductor photomultiplier tube (PMT) operated in two-photon absorption (TPA) regime.
The TPA detector measures the second order correlation function of the light field,
\begin{equation} \label{eq:G2}
G^{(2)} \equiv \langle E^{(-)}(t) E^{(-)}(t+\tau) E^{(+)}(t+\tau) E^{(+)}(t) \rangle ,
\end{equation}
where $E^{(-)}(t)$ is the negative frequency part of quantized EM field reaching the TPA detector at time $t$; $E^{(-)}(t+\tau)$ is the negative frequency part of quantized EM field reaching the TPA detector at time $t+\tau$  \cite{2001Optical}. $E^{(-)}(t)=E_1^{(-)}(t)+E_2^{(-)}(t)$ signifies that each $E$ field in Eq.~(\ref{eq:G2}) comes from both arm $1$ and $2$ of the MI.

\begin{figure}
\centering
\includegraphics[totalheight=2.8in]{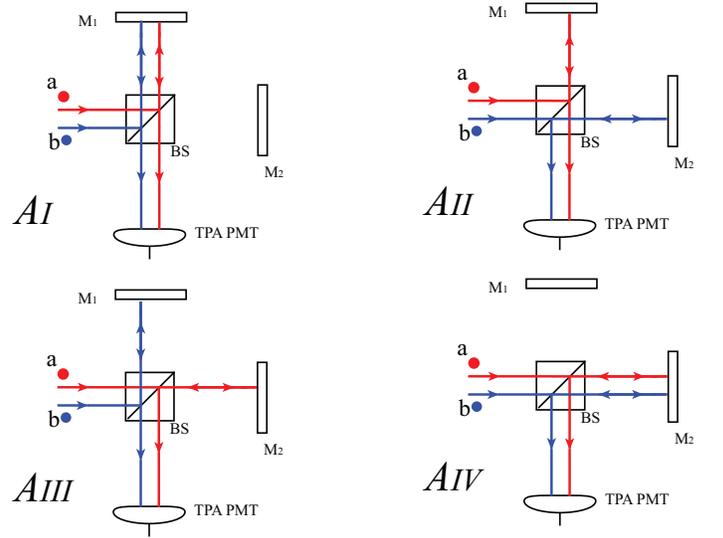}% Here is how to import EPS art
\dimendef\prevdepth=0
\caption{\label{fig:epsart} In the scalar model, there are four paths (probability amplitudes), $A_{I}$, $A_{II}$, $A_{III}$ and $A_{IV}$, to trigger an TPA detection event. In the vector model, we can manipulate the coefficients before each probability amplitudes by applying different polarizers combinations.}
\label{fig:4paths}
\end{figure}

From the quantum mechanical point of view, the signal of TPA detector in Eq.~(\ref{eq:G2}) can be calculated using the coherent superposition of four different and indistinguishable probability amplitudes.
Assuming the light is at single photon level, Eq.~(\ref{eq:G2}) can be written as \cite{2011An},
\begin{equation} \label{eq:G2-1}
G^{(2)} =|\langle 0| E_2^{(+)}(t+\tau) E_1^{(+)}(t) |1_a 1_b\rangle |^2,
\end{equation}
where $|1_a 1_b\rangle$ stands for the state of two photons $a$ and $b$; $E_1^{(+)}(t)$ and $E_2^{(+)}(t+\tau)$ signify $E$ fields come from arm $1$ and $2$ respectively. As shown in Fig.~\ref{fig:4paths}, there are four probability amplitudes involved in  Eq.~(\ref{eq:G2-1}) which are $A_{I}=A\substack{a\to1 \\ b\to1}$ , $A_{II}=A\substack{a\to1 \\ b\to2}$, $A_{III}=A\substack{a\to2 \\ b\to1}$ and $A_{IV}=A\substack{a\to2 \\ b\to2}$ from which we have,
\begin{equation} \label{eq:1}
G^{(2)}=|A_I+A_{II}+A_{III}+A_{IV}|^2,
\end{equation}
where $A_I$ to $A_{IV}$ are four probability amplitudes shown in Fig.~\ref{fig:4paths} \cite{tang2018measuring}.
The expansion of Eq.~(\ref{eq:G2}) has $16$ terms without taking polarizations into consideration. In general, each term has the form of $\langle E_{ai}^{(-)} E_{bj}^{(-)} E_{bl}^{(+)}  E_{ak}^{(+)} \rangle $ where $i,j,k,l=1,2$ stand for through which arms photons pass. For example, 
\begin{equation} \label{eq:G2-2}
A_{III}^* A_{II}=\langle E_{a2}^{(-)} (t+\tau)E_{b1}^{(-)}(t) E_{b2}^{(+)} (t+\tau) E_{a1}^{(+)}(t) \rangle.
\end{equation}

In Ref.~ \cite{tang2018measuring} a theoretical model based on the Feynman's path-integral and two-photon interference theory was developed to describe the Hanbury-Brown and Twiss effect (HBT) of multi-spatial-mode thermal light at ultrashort timescale by two-photon absorption. The theory is applied to interpret experimental results and shows that the output of the TPA detector is composed by four components which come from interference between different two-photon probability amplitudes.  In brief, the expansion of Eq.~(\ref{eq:1}) is comprised of four parts: the constant background which comes from $|A_I|^2+|A_{II}|^2$, the HBT term which come from $|A_{II}+A_{III}|^2$, the oscillation part with frequency $\omega$  and the oscillation part with frequency $2 \omega$.

If polarizations are taken into consideration, there are
\begin{eqnarray} \label{eq:polar-1}
E_1^{(+)}(t)& =& E_{1x}^{(+)}(t)+E_{1y}^{(+)}(t)\nonumber \\
E_2^{(+)}(t+\tau)& =& E_{2x}^{(+)}(t+\tau)+E_{2y}^{(+)}(t+\tau)  ,
\end{eqnarray}
where $E_{1x}^{(+)}(t)$ stands for the positive frequency part of $E$ field of $x$ polarization from channel $1$ which arrives the detector at time $t$, others terms have similar meanings. Combining  Eq.~(\ref{eq:G2-1}) and Eq.~(\ref{eq:polar-1}) we can see that each $16$ terms in Eq.~(\ref{eq:G2}) has the form of 
\begin{equation} \label{eq:G2-3}
\langle (E_{aix}^{(-)}+E_{aiy}^{(-)}) (E_{bjx}^{(-)}+E_{bjy}^{(-)}) (E_{blx}^{(+)}+E_{bly}^{(+)})  (E_{akx}^{(+)}+E_{aky}^{(+)}) \rangle  
\end{equation}
where $i,j,k,l=1,2$ stand for through which arms photons pass and $x,y$ stands for the polarizations. For example,  Eq.~(\ref{eq:G2-2}) changes into,
\begin{eqnarray} \label{eq:G2-4}
A_{III}^* A_{II}=\langle (E_{a2x}^{(-)} +E_{a2y}^{(-)})(E_{b1x}^{(-)}+E_{b1y}^{(-)})\nonumber \\(E_{b2x}^{(+)}+E_{b2y}^{(+)})  (E_{a1x}^{(+)}+E_{a1y}^{(+)}) \rangle,
\end{eqnarray}
in which there are $16$ terms after expansion. There are total $256$ terms in Eq.~(\ref{eq:G2}) after expansion.

However, not all the terms survive the expectation valuation $\langle \ldots \rangle$ in all $256$ terms because in general photon $a$ and $b$  from different polarization mode have different initial phases for chaotic light. The two-photon state of photons $a$ and $b$ can be written as,
\begin{equation} \label{eq:initial-phase}
|1_a 1_b\rangle=|1_a\rangle e^{i\delta_a}\otimes |1_b\rangle e^{i\delta_b},
\end{equation}
where $\delta_a$ and $\delta_b$ are random phases of photons $a$ and $b$ due to random excitations times of atoms respectively\cite{1997Quantum}. For photons from the same polarization, for example both photons come from $x$ polarization, we have $\langle e^{i (\delta_a -\delta_b)} \rangle=1$ which means that the initial phases of photons from the same polarization mode are completely correlated. If two photons are from orthogonal polarization, we have  $\langle e^{i (\delta_a -\delta_b)} \rangle=0$ which means that the initial phases of photons from the orthogonal polarizations are completely uncorrelated. 

Under this assumption, only $6$ out of $16$ terms survive in every terms in the expansion of Eq.~(\ref{eq:G2-3}), for example the expansion of Eq.~(\ref{eq:G2-4}) is,
\begin{eqnarray} \label{eq:G2-5}
%\hspace*{-1cm}
&&A_{III}^* A_{II} \nonumber \\ &= &\langle E_{a2x}^{(-)} E_{b1x}^{(-)} E_{b2x}^{(+)}  E_{a1x}^{(+)}\rangle + \langle E_{a2y}^{(-)} E_{b1y}^{(-)} E_{b2y}^{(+)}  E_{a1y}^{(+)}\rangle\nonumber \\
&+&\langle E_{a2x}^{(-)} E_{b1y}^{(-)} E_{b2x}^{(+)}  E_{a1y}^{(+)}\rangle+\langle E_{a2y}^{(-)} E_{b1x}^{(-)} E_{b2y}^{(+)}  E_{a1x}^{(+)}\rangle\nonumber \\
&+&\langle E_{a2x}^{(-)} E_{b1y}^{(-)} E_{b2y}^{(+)}  E_{a1x}^{(+)}\rangle+\langle E_{a2y}^{(-)} E_{b1x}^{(-)} E_{b2x}^{(+)}  E_{a1y}^{(+)}\rangle,\nonumber \\
\end{eqnarray}
where only in  these $6$ terms initial phases would cancel each other and have non-zero values and other $10$ terms equal to zeros. Since polarization is an independent dimension to describe the $EM$ field as same as time and space, the Eq.~(\ref{eq:G2-3}) can be factorized into the product of polarizations part and temporal part (all the calculation is assumed to be done in the same spatial mode) and written as,
\begin{eqnarray} \label{eq:G2-6}
&&A_{III}^* A_{II} \nonumber \\  & = &[\langle E_{2x}^{(-)} E_{1x}^{(-)} E_{2x}^{(+)}  E_{1x}^{(+)}\rangle + \langle E_{2y}^{(-)} E_{1y}^{(-)} E_{2y}^{(+)}  E_{1y}^{(+}\rangle\nonumber \\
&+&\langle E_{2x}^{(-)} E_{1y}^{(-)} E_{2x}^{(+)}  E_{1y}^{(+)}\rangle+\langle E_{2y}^{(-)} E_{1x}^{(-)} E_{2y}^{(+)}  E_{1x}^{(+}\rangle\nonumber \\
&+&\langle E_{2x}^{(-)} E_{1y}^{(-)} E_{2y}^{(+)}  E_{1x}^{(+}\rangle+\langle E_{2y}^{(-)} E_{1x}^{(-)} E_{2x}^{(+)}  E_{1y}^{(+)}\rangle]  \nonumber \\
& \times & \langle E_{a2}^{(-)} E_{b1}^{(-)} E_{b2}^{(+)}  E_{a1}^{(+)}\rangle ,\nonumber \\
\end{eqnarray}
where $\langle E_{a2}^{(-)} E_{b1}^{(-)} E_{b2}^{(+)}  E_{a1}^{(+)}\rangle$ corresponds to the temporal interference term in the scalar model \cite{tang2018measuring} and the sum of $6$ terms in $[..]$ correspond to the polarization interference only found in the vector model. From Eq.~(\ref{eq:G2-6}) we can see that in the vector model polarizations determine the coefficients of interference terms in the scalar model. Since other $16$ terms in the expansion of Eq.~(\ref{eq:G2-1}) have the similar form as shown in Eq.~(\ref{eq:G2-6}), in the vector model we have,
\begin{equation} \label{eq:VSmodel}
V_{TPA}=C \otimes S_{TPA},
\end{equation}
where $V_{TPA}$ is the probability density matrix in vector model, $C$ is the coefficients matrix which will be defined lately, $\otimes$ stands for Hadamard product and $S_{TPA}$ is the probability density matrix derived in scalar model which is defined as \cite{tang2018measuring}, 

\begin{small}
\begin{equation} \label{eq:Smode}
\centering
%\hspace*{-1cm}
S_{TPA} =
\left[ \begin{array}{ccccc}
A_{I}^* A_{I}&A_{I}^* A_{II}&A_{I}^* A_{III} &A_{I}^* A_{IV}& \\
A_{II}^* A_{I}&A_{II}^* A_{II}&A_{II}^* A_{III}&A_{II}^* A_{IV}& \\
A_{III}^* A_{I}&A_{III}^* A_{II}&A_{III}^* A_{III} &A_{III}^* A_{IV}& \\
A_{IV}^* A_{I}&A_{IV}^* A_{II}&A_{IV}^* A_{III} &A_{IV}^* A_{IV} & \\
\end{array} \right],
\end{equation}
\end{small}
where terms like $A_{II}^* A_{III}$ now stand for the interference term in the scalar model in which only temporal interference is taken into consideration. The coefficients matrix $C$ is defined as,
%\begin{scriptsize}
\begin{equation} \label{eq:c-matrix}
\centering
%\hspace*{-1cm}
C =
\left[ \begin{array}{ccccc}
c_{1111}&c_{1112}&c_{1121} &c_{1122}& \\
c_{1211}&c_{1212}&c_{1221}&c_{1222}& \\
c_{2111}&c_{2112}&c_{2121} &c_{2122}& \\
c_{2211}&c_{2212}&c_{2221} &c_{2222} & \\
\end{array} \right],
\end{equation}
%\end{scriptsize}
where 
\begin{eqnarray} \label{eq:c-matrix-1}
\centering
%\hspace*{-1cm}
&& c_{ijkl} \nonumber \\ &=& \langle E_{ix}^{(-)} E_{jx}^{(-)} E_{lx}^{(+)}  E_{kx}^{(+)}\rangle + \langle E_{iy}^{(-)} E_{jy}^{(-)} E_{ly}^{(+)}  E_{ky}^{(+)}\rangle\nonumber \\
&+&\langle E_{ix}^{(-)} E_{jy}^{(-)} E_{lx}^{(+)}  E_{ky}^{(+)}\rangle+\langle E_{iy}^{(-)} E_{jx}^{(-)} E_{ly}^{(+)}  E_{kx}^{(+)}\rangle\nonumber \\
&+&\langle E_{ix}^{(-)} E_{jy}^{(-)} E_{ly}^{(+)}  E_{kx}^{(+)}\rangle+\langle E_{iy}^{(-)} E_{jx}^{(-)} E_{lx}^{(+)}  E_{ky}^{(+)}\rangle,\nonumber \\
\end{eqnarray}
where $i,j,k,l=1,2$ stand for through which arms photons pass and $x,y$ stand for polarizations.

To make the calculation easier we define a \emph{second-order covariance matrix} or \emph{two-photon covariance matrix} (TCM) $J^{(2)} $ since it describes the annihilation of two-photons with polarizations \cite{2001Optical},
\begin{equation} \label{eq:connection}
\centering
J^{(2)}(i,j,k,l) =
\left[ \begin{array}{ccccc}
{\color{red} J_{xxxx}} &J_{xxxy} &J_{xxyx}  &J_{xxyy} & \\
J_{xyxx} &{\color{red} J_{xyxy}} &{\color{red} J_{xyyx}}   &J_{xyyy} & \\
J_{yxxx} &{\color{red} J_{yxxy}}  &{\color{red} J_{yxyx}}  &J_{yxyy} & \\
J_{yyxx} &J_{yyxy} &J_{yyyx}  &{\color{red} J_{yyyy}}  & \\ 
\end{array} \right], 
\end{equation}
where $i,j,k,l=1,2$ have the same meanings defined in  Eq.~(\ref{eq:c-matrix}), $x,y$ stand for the polarization and the positions of subindexes of $J$ are define as: the first and the fourth indexes correspond the EM field of photon $a$  and the second and third indexes correspond the EM field of photon $b$. For example the element $J_{xyyx}(i,j,k,l)\equiv
\langle E_{aix}^{(-)} E_{bjy}^{(-)} E_{bly}^{(+)} E_{akx}^{(+)} \rangle$ stands for the second order correlation function of $E$ fields of $x$ polarization of photon $a$ through path $i$, $E$ fields of $y$ polarization of photon $b$ through path $j$, $E$ fields of $y$ polarization of photon $b$ through path $l$ and $E$ fields of $x$ polarization of photon $a$ through path $k$. The connect between  Eq.~(\ref{eq:c-matrix}) and  Eq.~(\ref{eq:connection}) is,
\begin{eqnarray} \label{eq:connection-1}
c_{ijkl} & = &J^{(2)}[1,1]+J^{(2)}[2,2]+J^{(2)}[2,3]\nonumber\\
& + & J^{(2)}[3,2]+J^{(2)}[3,3]+J^{(2)}[4,4],
\end{eqnarray}
where in Eq.~(\ref{eq:connection}) only the $6$ red terms (color online) are not zero and contribute to $G^{(2)}$.

One of the advantages of defining the two-photon covariance matrix is that the setup shown in Fig.~\ref{fig:setup} is a linear system and the EM field operators and the two-photon coherence matrix at the TPA detector relate to those at the input of the MI by a linear transformation matrix which is determined by the experimental setups \cite{2001Optical}. The polarized MI we studied is comprised of polarizers, non-polarized beams splitter and mirrors.  The connection between the TCM $J^{(2)}$ at the TPA detector and the TCM  $J_0^{(2)}$ at the input of the MI is \cite{2001Optical}
\begin{equation} \label{eq:tm}
J^{(2)}= (T_1 T_2 \ldots T_n)^ \dagger J_0^{(2)} (T_1 T_2 \ldots T_n),
\end{equation}
where $T_1 T_2 \ldots T_n$ stands for the cascade transformation matrix for the  MI. Once the two-photon coherence matrix is determined the $G^{(2)}$ function of the polarized MI could be calculated using Eq.~(\ref{eq:VSmodel}) in which the coefficients matrix $C$ is calculated using Eq.~(\ref{eq:connection-1}).

\section{\label{sec:simul}Simulations}
In this section, we will employ the method above to study two-photon interference in different schemes and show how to manipulate the interference. In simulations, all the figure plot the normalized second order correlation functions  $g^{(2)}=\frac{G^{(2)}}{\langle E_1^{(-)}E_1^{(+)} \rangle \langle E_2^{(-)}E_2^{(+)} \rangle}$\cite{2001Optical}.

\subsection{\label{subsec:unpolar}Unpolarized chaotic light as input}
\begin{figure}
\centering
\hspace*{-1cm}
\includegraphics[totalheight=1.75in]{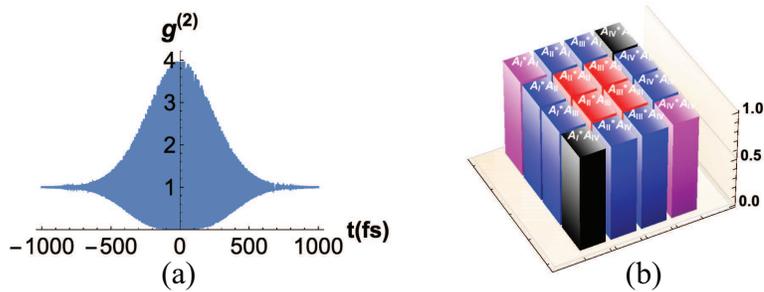}
%\caption{a}
%\includegraphics[totalheight=1in]{no-polarizers-2}% Here is how to import EPS art
\dimendef\prevdepth=0
\caption{\label{fig:epsart} (a) is the simulated output of the TPA detector without polarizer in the MI. (b) is its visualized probability density matrix. The two megenta bars stand for the constant background. The eight blue bars stand for oscillation patterns with $\omega$ frequency. The four red bars stand for HBT effect. The two black bars stand for oscillation patterns with $2\omega$ frequency. Each bar is from different probability amplitudes interference which is labelled on the top of the bar.}
\label{fig:no-polarizers}
\end{figure}

We start with the unpolarized chaotic light. In this case, polarizers are absent in the MI shown in Fig.~\ref{fig:setup}. Without polarizers involved,  there are four kinds of  interference patterns in the outcomes of the TPA detection as mentioned in Sec.~\ref{sec:theory}. They are constant background, HBT effect, the oscillation pattern with frequency $\omega$ and the oscillation pattern with frequency $2 \omega$ respectively \cite{tang2018measuring}. The output of the TPA detector is the sum of each elements of probability density matrix $S_{TPA}$ as shown in Eq.~(\ref{eq:G2-1}). All the four different components are mixed together and shown in Fig.~\ref{fig:no-polarizers}(a). 

To have a better understanding of the structure of interference patterns, the probability density matrix $S_{TPA}$ are visualized by using a $3D$ barchart in which the height of bars are proportional to their relative probabilities of each element. In $S_{TPA}$, many terms are complex number and their real parts are taken as their relative probabilities. In the barchart, the constant background part which corresponds  to the two-photon probability that two-photons come from either arm $1$ or $2$ is visualized by two magenta bars in Fig.~\ref{fig:no-polarizers}(b). This component does not change with the relative arrival time difference $\tau$ between two photons. The second component corresponds to the well known HBT effect. It  describes that photons $a$ and $b$ trigger the TPA detector in two different ways : photon $a$ comes from arm $1$ and photon $b$ comes from arm $2$ which corresponds to two-photon amplitude $A_{II}$ ;  photon $a$ comes from arm $2$ and photon $b$ comes from arm $1$ which corresponds to two-photon amplitude $A_{III}$ as shown in Fig.~\ref{fig:4paths}. The probability of HBT effect is $|A_{II}+A_{III}|^2$. This component is visualized by four red bars in Fig.~\ref{fig:no-polarizers}(b). The third component of TPA detection can be factorized into the product of intensity and first oder interference and it is visualized by eight blue bars in Fig.~\ref{fig:no-polarizers}(b). The fourth part is interesting because it stands for that photon $a$ and $b$ interference with themselves as one entity. In the expansion of Eq.~(\ref{eq:G2-1}) it is signified by the term of $A^*_IA_{IV}+A_IA^*_{IV}$, two photons can come from either arm $1$ or $2$ as one entity, the two probability amplitudes interfere with each other and leads to sub-wavelength effect. The fourth component is visualized by two black bars in Fig.~\ref{fig:no-polarizers}(b). In an ordinary HBT interferometer, only the HBT effect is measured because other parts are ruled out by the detection scheme of an ordinary HBT interferometer \cite{brown1956correlation}. However, in a MI with a point TPA detector all these four kinds of TPA events exit and mix together.  In previous research, people usually concentrated on the HBT effect part plus the inevitable constant background which are signified by four red bars and two magenta bars in the probability matrix  and filter out the third and fourth parts which is signified by eight blue bars and two black bars in the probability matrix \cite{boitier2009measuring,shevchenko2017polarization,2018Recovering,2015Ultrabroadband}. In this paper, we will take every parts into consideration and find the method to manipulate the interference process using two-photon interference theory which leads to interesting results. 

When the input of the MI is unpolarized chaotic light the outcome of the TPA detector is shown in Fig.~\ref{fig:no-polarizers}(a) which was measured in almost every previous researches using similar detection schemes \cite{boitier2009measuring,shevchenko2017polarization,2018Recovering,2015Ultrabroadband}. It is proportional to the sum of $16$ different probabilities to trigger a TPA event which are shown in Fig.~\ref{fig:no-polarizers}(b) and we can see that each $16$ probabilities are equal. The sum of all these probabilities lead to the $g^{(2)}$ function as shown in Eq.~(\ref{eq:1}).

\begin{figure}[h]
\centering
\hspace*{-6cm}
\subfigure{
\begin{minipage}[t]{0.25\linewidth}
\centering
\includegraphics[totalheight=1.5in]{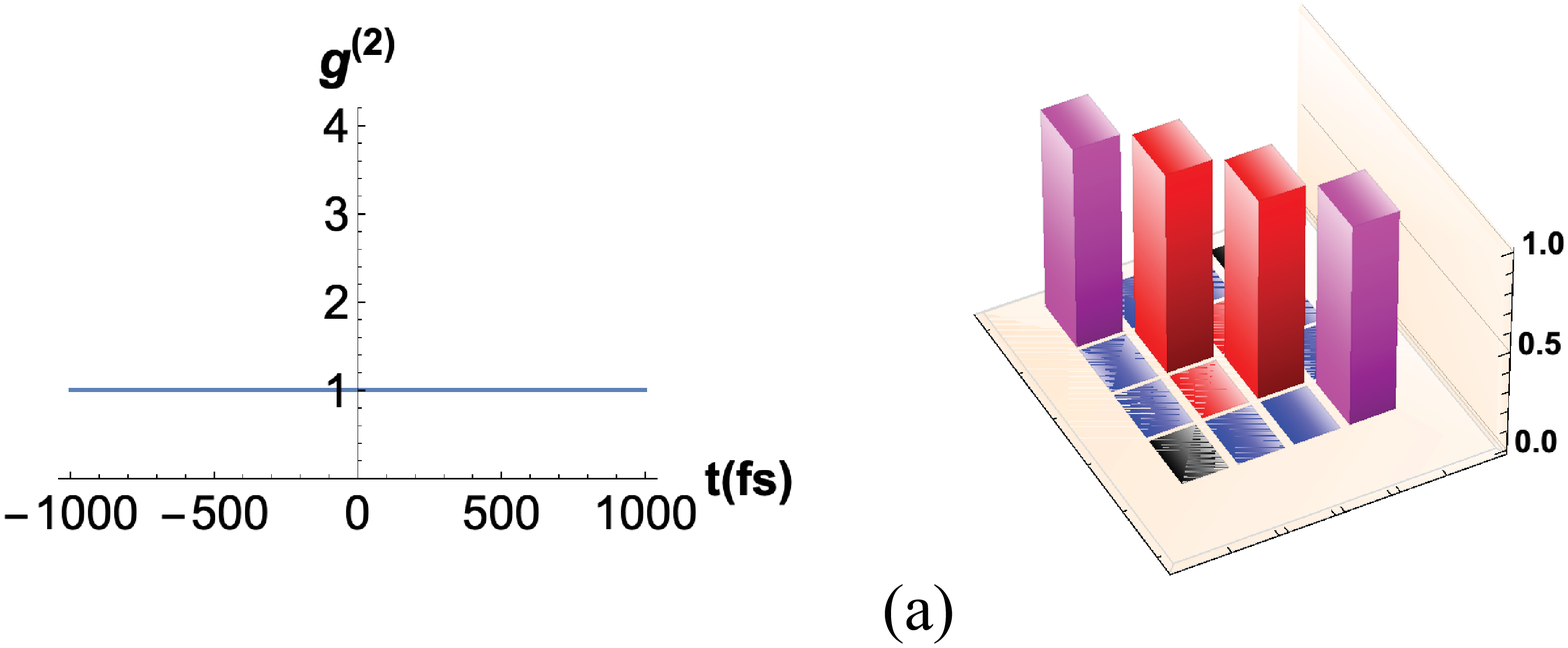}
%\caption{fig1}
\end{minipage}%
}%

\subfigure{
\begin{minipage}[t]{0.25\linewidth}
\centering
\hspace*{-3cm}
\includegraphics[totalheight=1.5in]{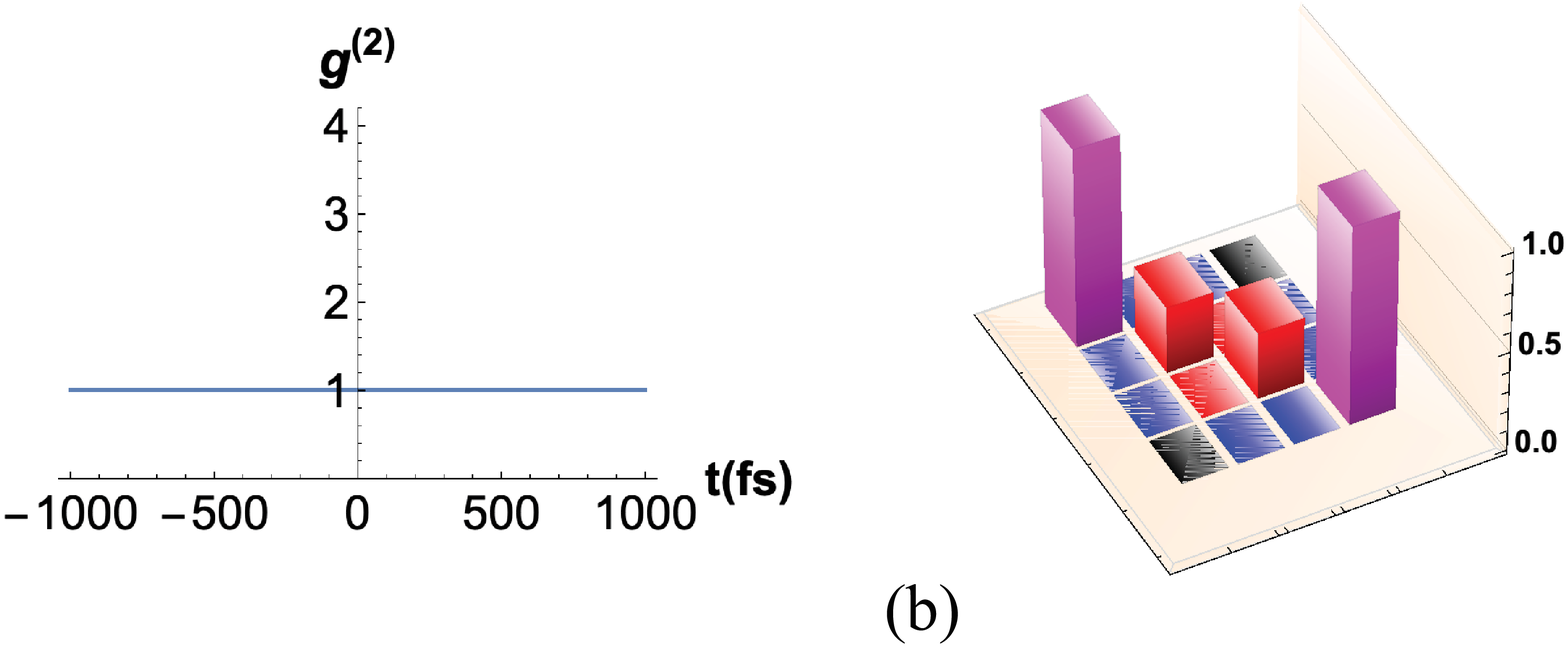}
%\caption{fig2}
\end{minipage}%
}%
\centering
\caption{ Comparison between  $0^{\circ}/90^{\circ}$ and $45^{\circ}/135^{\circ}$ schemes. The output of $0^{\circ}/90^{\circ}$ scheme is shown in (a) and that of $45^{\circ}/135^{\circ}$ scheme is shown in (b). The have the same $g^{(2)}(t)$ function but they have different components distribution in the function. In $45^{\circ}/135^{\circ}$ scheme, the two-photon is more likely coming from the same arm of the interferometer.}
\label{fig:0-90vs45-45}
\end{figure}

Next we simulate the two-photon interference of photons from orthogonal polarizations in two different cases. Two polarizers, $P_1$ and $P_2$, which are orthogonal to each other are inserted into arm $1$ and $2$. In the first case polarizer $P_1$ is set to $0$ in arm $1$ and polarizer $P_2$ is set to $\frac{\pi}{2}$ in arm $2$ as shown in Fig.~\ref{fig:setup}, the output of the TPA detector is shown in Fig.~\ref{fig:0-90vs45-45}(a). 

In the second case $P_1$ is set to $\frac{\pi}{4}$ in arm $1$ and $P_2$ is set to $-\frac{\pi}{4}$ in arm $2$, the output of the TPA detector is shown in Fig.~\ref{fig:0-90vs45-45}(b). 

Comparing  (a) with (b) in Fig.~\ref{fig:0-90vs45-45}, we can see that they are both flat in the center of $g^{(2)}$ function, $g^{(2)}(0)=1$. This means that the two-photon interference generates no bunching effect if photon $a$ and $b$ come  from orthogonal polarization modes (in (a) they are set to $0^{\circ}$/ $90^{\circ}$ and in (b) they are set to $45^{\circ} $/$135^{\circ}$). This can be explained as that photons from orthogonal polarization modes has uncorrelated initial phases. The terms which lead to bunching effect $A_{III}^* A_{II} +A_{III} A_{II}^*=0$. However, no bunching effect does not mean no two-photon interference. The two-photon interference leads to the possibility distribution of triggering a TPA events different in two cases. There are four possibility to trigger a TPA event in both two cases: photon $a$ and $b$ can both come from arm $1$ or $2$ which are $|A_{a1b1}|^2$ and  $|A_{a2b2}|^2$ respectively and correspond to two magenta columns in Fig.~\ref{fig:0-90vs45-45}; photon $a$ from arm $1$ and photon $b$ from arm $2$ which corresponds to $|A_{a1b2}|^2$;  photon $a$ from arm $2$ and photon $b$ from arm $1$ which corresponds to $|A_{a2b1}|^2$, the last two possibility correspond to two red columns in Fig.~\ref{fig:0-90vs45-45}. 

We notice that in the two schemes the possibilities distribution for a TPA detection is different. For $0^{\circ}/90^{\circ}$ scheme, all the possibilities is the same and equal to $\frac{1}{4}$. However, for  $45^{\circ}/135^{\circ}$ scheme, the possibility for both photons come from the same arm (either arm $1$ or $2$) is $\frac{3}{8}$; the possibility for both photons come from different arm is $\frac{1}{8}$. This result is non-intuitive. Even the  $g^{(2)}$ function is the same, the contributions from four possibilities are different. 

\begin{figure}
\centering
\includegraphics[totalheight=1.5in]{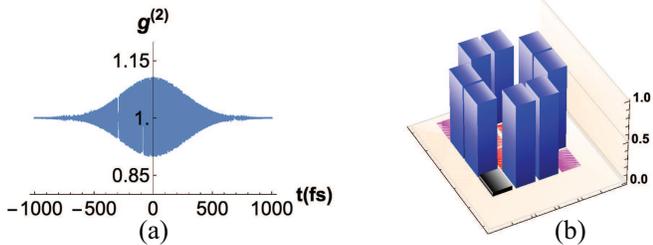}
%\caption{a}
%\includegraphics[totalheight=1in]{no-polarizers-2}% Here is how to import EPS art
\caption{\label{fig:epsart} In the $45^{\circ}/135^{\circ}$ scheme  if a polarizer $P_3$ which is set to $0^{\circ}$ is added in front of  the TPA detector, and set one of the polarizer say $P_1$ deviate from the $45^{\circ}$ a few degree, $\omega$  oscillation part of two-photon interference will dominate comparing with the $2\omega$ oscillation part. (a) shows the output of the TPA detector. (b) shows the probability density matrix in which the constant backgrounds are removed.}
\label{fig:omega-dominating}
\end{figure}

As shown in Eq.~(\ref{eq:VSmodel}), polarizations can be used to manipulate the two-photon interference in the MI. In the $45^{\circ}/135^{\circ}$ scheme  if a polarizer $P_3$ which is set to $0^{\circ}$ is added in front of  the TPA detector, and set one of the polarizer say $P_1$ deviate from the $45^{\circ}$ a few degree, $\omega$  oscillation part of two-photon interference will dominate comparing with the $2\omega$ oscillation part. It is shown in Fig.~\ref{fig:omega-dominating}
in which the probability term of $|A_{I}|^2$, $|A_{II}|^2$, $|A_{III}|^2$ and $|A_{IV}|^2$ are removed to make a comparison between only $\omega$ and $2\omega$ terms. In the next subsection there is a detection scheme in which the $\omega$ terms are removed and only $2\omega$ terms is detected which leads to sub-wavelength effect. 
\subsection{\label{subsec:polar}Polarized chaotic light and its sub-wavelength effect}

Now we put a linear polarizer $P_0$ in front of the beamsplitter as shown in Fig.~\ref{fig:setup}, it turns the unpolarized chaotic light into linear polarized light before into the MI. When there is no polarizer in both arms the $g^{(2)}$ function and two-photon detection possibility matrix are the same as those in unpolarized light case as show in Fig.~\ref{fig:no-polarizers}. 

\begin{figure}
\centering
\includegraphics[totalheight=1.5in]{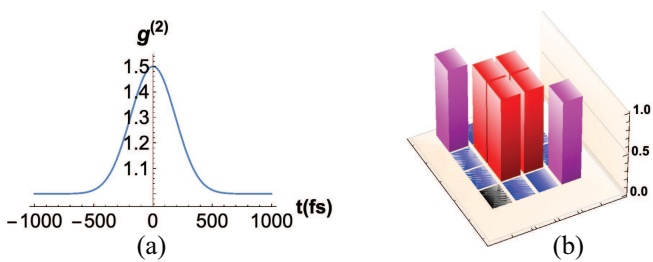}
%\caption{a}
%\includegraphics[totalheight=1in]{no-polarizers-2}% Here is how to import EPS art
\caption{\label{fig:epsart} When the input of linear polarized chaotic light is set to $45^{\circ}$ and polarizers are set to $0^{\circ}$ and $90^{\circ}$ in two arms respectively, all the oscillation patterns are removed. Only HBT effect and background are observed.}
\label{fig:45-0-90}
\end{figure}

If we set the polarizer $P_0$ to $45^{\circ}$, $P_1$ to $0^{\circ}$ and $P_2$ to $90^{\circ}$ the $g^{(2)}$ function and its two-photon detection possibility matrix are shown in Fig.~\ref{fig:45-0-90}. We can see that there is bunching effect but there is no $\omega$ and $2\omega$ oscillation terms. The reason is that from the point of view of quantum interference the TPA detector can \emph{in principle} identify from which arms(paths) photons come from because of the two polarizers in arms $1$ and $2$. Since the $which path$ information is known, there is no corresponding two-photon interference. However, the probability amplitudes $A_{II}$ and $A_{III}$ are stilled indistinguishable and the interference between them leads to the HBT effect as shown in four red columns in Fig.~\ref{fig:45-0-90}.

\begin{figure}
\centering
\includegraphics[totalheight=1.5in]{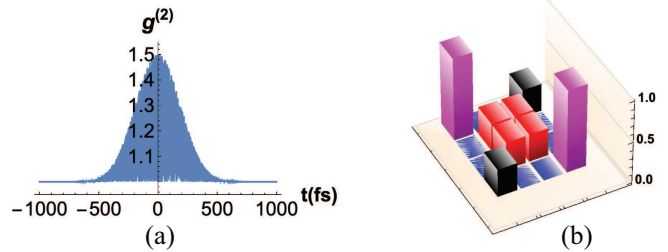}
%\caption{a}
%\includegraphics[totalheight=1in]{no-polarizers-2}% Here is how to import EPS art
\caption{\label{fig:epsart} When the input of linear polarized chaotic light is set to $0{\circ}$ and polarizers are set to $45^{\circ}$ and $135^{\circ}$ in two arms respectively, the oscillation pattern with $\omega$ frequency is removed. The oscillation pattern with $2 \omega$ frequency remains. As shown in probability density matrix($2$ black bars), photon $a$ and $b$ interferes with themselves as one entity which leads to the sub-wavelength effect. HBT effect and background also remain.}
\label{fig:0-45-135}
\end{figure}

If we set the polarizer $P_0$ to $0^{\circ}$, $P_1$ to $45^{\circ}$ and $P_2$ to $135^{\circ}$ the situation is more interesting. The $g^{(2)}$ function and its two-photon detection possibility matrix are shown in Fig.~\ref{fig:0-45-135}. We can see that there is bunching effect and no $\omega$  oscillation terms. Surprisingly there is $2\omega$ oscillation terms as shown in two black columns in the figure. From the point view of quantum optics, photon $a$ and $b$ form one entity which interferes with itself. The interference patterns have $2\omega$ frequency of oscillation. This is a sub-wavelength effect. The corresponding experimental phenomenon has been observed and the details are reported in another paper \cite{luo2021observing}. 

\begin{figure}[h]
\centering
\hspace*{-8cm}
\subfigure{
\begin{minipage}[t]{0.25\linewidth}
\centering
\includegraphics[width=3.5in]{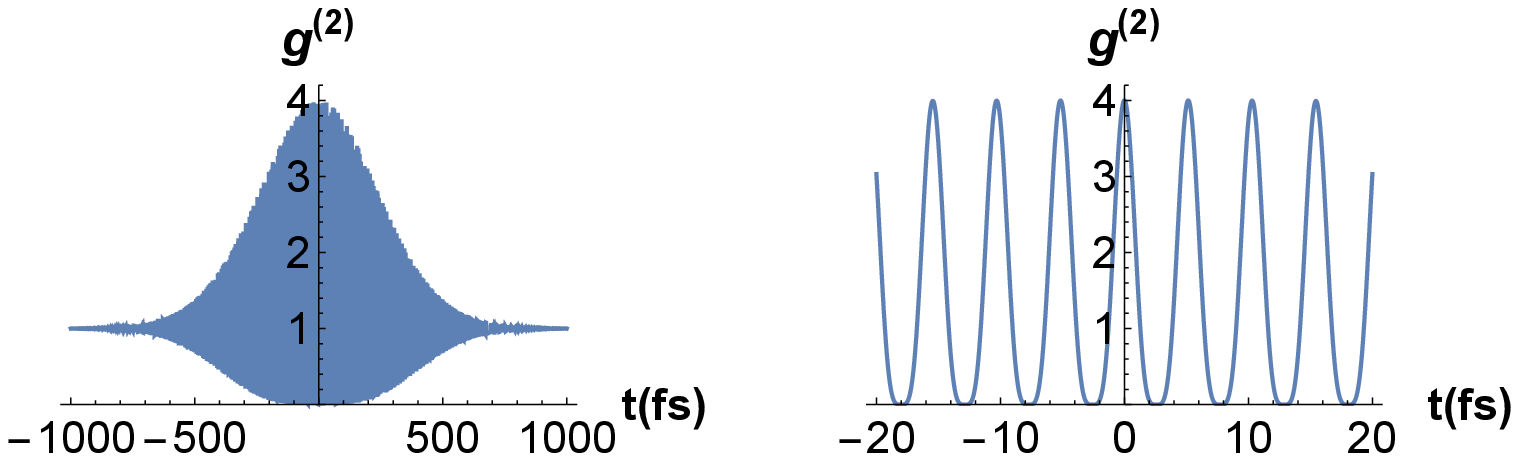}
%\caption{fig1}
\end{minipage}%
}%

\subfigure{
\begin{minipage}[t]{0.25\linewidth}
\centering
\hspace*{-3.9cm}
\includegraphics[width=3.5in]{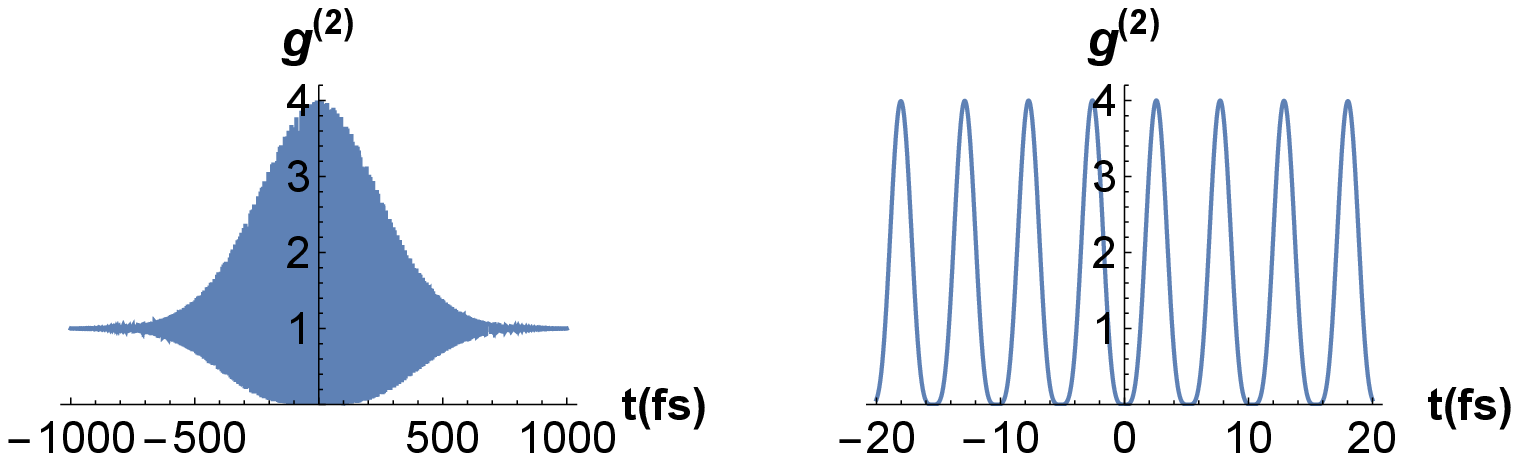}
%\caption{fig2}
\end{minipage}%
}%

\centering
\caption{ The polarizer $P_3$ in front of the TPA detector can erase the $which path$ information. By controlling the relative angle between $P_0$ and $P_3$ ($0^{\circ}$ for the top and $90^{\circ}$ for the bottom of the figure) we can choose to observe maximum or minimum of the interference pattern at the balance position of the MI.}
\label{fig:45vs135}
\end{figure}

With another polarizer $P_3$ is put before the detector, it could acts a \textit{which path} information eraser. For example, when polarizer $P_0$ is set to $45^{\circ}$, $P_1$ to $0^{\circ}$ and $P_2$ to $90^{\circ}$  and the output is shown in Fig.~\ref{fig:45-0-90}. With polarizer $P_3$ is set to $45^{\circ}$ before the TPA detector, the output of the detector and the probability matrix is resumed as same as those shown in Fig.~\ref{fig:no-polarizers} because the $which path$ information is erased by polarizer $P_3$ and interference terms leads to oscillation is not zero anymore. If $P_3$ is set to $135^{\circ}$ instead of $45^{\circ}$ the situation is slightly different: the probability matrix is the same but the $g^{(2)}(0)$ change from maximum to minimum as shown in Fig.~\ref{fig:45vs135}.

\section{Discussion}

First, we notice that  the outputs of TPA detection are slightly different under two different schemes as shown in  Fig.~\ref{fig:0-90vs45-45}. In both $0^{\circ}/90^{\circ}$ and $45^{\circ}/135^{\circ}$ schemes they both have flat $g^{(2)}$ functions and are comprised of four possibilities $|A_{I}|^2$, $|A_{II}|^2$, $|A_{III}|^2$ and $|A_{IV}|^2$. They are different in the percentages of contributions from the four possibilities. In $0^{\circ}/90^{\circ}$ scheme, each of the four possibilities contributes $\frac{1}{4}$ to $g^{(2)}$. However, in $45^{\circ}/135^{\circ}$ scheme, each of $|A_{I}|^2$ and  $|A_{IV}|^2$ contributes $\frac{3}{8}$ to $g^{(2)}$; each of $|A_{II}|^2$ and  $|A_{III}|^2$ contributes $\frac{1}{8}$ to $g^{(2)}$. The reason for the difference lies in the mirror reflection of the BS. In $0^{\circ}/90^{\circ}$ scheme, the reflection of the BS does not change the polarizations of light. In $45^{\circ}/135^{\circ}$ scheme, however, there is a mirror reflection from the BS which changes the left and right to make the polarization of $45^{\circ}/135^{\circ}$ switch to $135^{\circ}/45^{\circ}$. So in order to make the a $45^{\circ}/135^{\circ}$ detection scheme as shown in Fig.~\ref{fig:0-90vs45-45}(b) we need to set both $P_1$ and $P_2$ to $135^{\circ}$ because the $45^{\circ}$ polarization of chaotic light will enter into arm $1$ in the angle of $135^{\circ}$ because of the reflection of the BS.  The reflection of the BS leads to the difference between the two-photon coherence covariance (TCM) of $0^{\circ}/90^{\circ}$ scheme and that of $45^{\circ}/135^{\circ}$ scheme and at last the differences between their TPA detection probability density matrixes.

The above simulated results can be verified experimentally. In both schemes, we can measure their total TPA detection rates and assume they are all equal to $1$. Then we can block one arm, say arm $2$, and only two-photon probability $|A_{I}|^2$ is not blocked. In $0^{\circ}/90^{\circ}$ scheme, the TPA detection rates should drop to $\frac{1}{4}$.  In $45^{\circ}/135^{\circ}$ scheme, the TPA detection rates should be $\frac{3}{8}$, slightly higher than that in $0^{\circ}/90^{\circ}$ scheme.

The reflection of the BS is also the reason for the difference in polarized chaotic light in Sec.~\ref{subsec:polar}. In $45^{\circ}/0^{\circ}/90^{\circ}$ scheme, all the $\omega$ and $2\omega$ oscillation parts are removed, only the HBT effect and constant background are left. On the other hand, in $0^{\circ}/45^{\circ}/135^{\circ}$ scheme only  the $\omega$ oscillation part is removed and other than the HBT effect and constant background the  $2\omega$  part also exists. In the point of view of quantum mechanics, the $2\omega$ oscillation part is a sub-wavelength effect from which an entity comprised of photons $a$ and $b$ interferes with itself \cite{1995Photonic,1999Measurement,2001Two}. The momentum of the entity is twice that of a single photon and the De Broglie wavelength is half of a single photon. The sub-wavelength effect can be used to increase the resolution of imaging or quantum lithography \cite{2001Two}. The sub-wavelength effect predicted by the vector model has been observed in our following experiments and reported in another paper \cite{luo2021observing}. 

In Sec.~\ref{subsec:polar}  it is found that by controlling the relative angle between polarizer $P_0$ and $P_3$ the value of $g^{(2)}(0)$ can be manipulated as shown in Fig.~\ref{fig:45vs135}. When $P_0$ is set to parallel to $P_3$ $g^{(2)}(0)$ reaches its maximum value and when $P_0$ is set to orthogonal to $P_3$ $g^{(2)}(0)$ reaches its minimum value. This scheme could be applied in our previously proposed weak signal detection MI by exploring super-bunching effect of chaotic light \cite{2020Two}. In a LIGO-like weak signal detection interferometer, to have higher sensitivity and save energy the detector is made to observe the dark fringe \cite{Harry2010Advanced}. In our proposed new weak-signal detection scheme  dark fringe can be manipulated by adjusting the relative angle between polarizers $P_0$ and $P_3$.
\section{Conclusion}

In this paper, a vector model is developed to theoretically describe the two-photon interference phenomenon of chaotic light in a MI with polarizers. The model is developed by using two-photon interference and Feynman path integral theory. The model shows that the polarization as an independent dimension in phase space can act as a switch to manipulate the two-photon interference in the MI. The components of two-photon interference patterns which are mixed together in previous studies can now be picked out one by one by adjusting polarizers in the MI. The vector model could help us in further study in a cascaded MI which explores super-bunching effect of chaotic light to increase the sensitivity on weak signal detection \cite{2020Two}. It may help us to design a new type of weak signal (including gravitational wave) detection setup with higher sensitivity.    

\begin{acknowledgments}
This work was supported by Shaanxi Key Research and Development Project (Grant No. 2019ZDLGY09-09); National Natural Science Foundation of China (Grant No. 61901353); National Nature Science Foundation of China  (Grant No. 12074307); Key Innovation Team of Shaanxi Province (Grant No. 2018TD-024) and 111 Project of China (Grant No.B14040).
\end{acknowledgments}

\appendix

\bibliographystyle{unsrt}
\bibliography{pmi-2}% Produces the bibliography via BibTeX.

\end{document}